\documentstyle[preprint,aps,psfig]{revtex}

\begin{document}

\title{Two Landmarks in Polymer Physics: The Edwards Model and de Gennes
Observation}

\author{D. Thirumalai}

\address{Institute for Physical Science and Technology
University of Maryland, College Park, MD 20742}

\maketitle

\begin{abstract}

The impact of two landmark papers by Edwards and de Gennes on the field of
polymer physics is highlighted.

\end{abstract}

\vspace{0.5in}


\vspace{0.5in}

A major advance in the physics of polymers occurred in 1949 when Flory [1]
provided a simple but profound argument for the swelling (compared to the
ideal chain size) of flexible polymer
chains due to excluded volume interactions. In essence, the Flory result for
the dependence of the radius of gyration R$_g$ on the degree of
polymerization N is obtained by minimizing the elastic energy (due to chain
connectivity) and the repulsive energy arising from volume excluded by a
given monomer for all other monomers. The resulting prediction for the
exponent $\nu $ defined by R$_g\approx aN^\nu $ is remarkably accurate in
all space dimensions, d. For all practical purposes the Flory result for $
\nu =\frac 3{d+2}$ [2a] may be considered exact [2b]. Similarly, the 
Flory argument is
also found to be nearly exact for describing sizes of D dimensional objects
embedded in d spatial dimensions,
such as tethered membranes with D = 2 [3]. For polymers D = 1. 

The fundamental understanding of the reasons for the success of the theory
due to Flory is still lacking. In an attempt to derive the Flory exponent
Edwards proposed a model for polymers that bears his name in 1965 [4]. This
paper brought to bear, for the first time, methods of functional integrals and
many body theory to bear on problems in polymer physics. Edwards proposed a
very simple form for the short range repulsive potential describing the
interactions between the monomers. He suggested that replacing the actual
potential by $\upsilon \delta [r(s)-r(s^{\prime })]$ where $\upsilon $ is
the strength of the excluded volume interactions, $r(s)$ is a path of the
polymer chain and $s$ and $s^{\prime }$ are the positions of two monomers
along the positions of the chain. The use of the delta function pseudo
potential should not (see below) affect the long wavelength properties of the
polymer chain. With this replacement Edwards formally showed that polymer
statistics boils down to summing over all possible paths weighted by the
Hamiltonian given by the sum of the ''kinetic energy'' (representing chain
connectivity) and the pseudo potential. The resulting path integral is
non-Markovian which is a reflection of the nature of the excluded volume
interactions. The formal analogy to path integral allowed the use of many
approximations devised in the context of quantum mechanics to problems in
polymers. 

Several studies utilizing the Edwards model for polymers followed [5]. In
addition using
enumerations of self-avoiding 
walks using lattice models [6], and
through the ingenious use of exact relations for Ising models [7] many new
results for polymer statistics were obtained.  However, 
an understanding of the varied universal
behavior of polymer solutions was lacking. This state of affairs in polymer
physics was to change dramatically after the profound discovery by de Gennes
who showed a connection between polymer statistics and phase transitions in
1972 [8]. This short and lucid paper followed right at the 
heels of the discovery of the
renormalization group in the context of second order phase transitions. de
Gennes showed that the n-vector magnetic spin problem with n = 0 is
equivalent to the excluded volume problem considered by Flory [1] and formalized
in terms of path integral methods by Edwards [4]. The connection between
the excluded volume problem and phase transitions also clarified the reasons
for the independence of the values of $\nu$ on the details of the interaction
potentials as long as they are short ranged. This, in retrospect, justified
the Edwards choice of delta function interaction between two monomer segments.
With the profound  observation by de Gennes
the entire machinery developed for understanding critical phenomenon could
be imported to obtain a vast number of new results. Thus, the concept of
scaling was born in polymer physics and continues to dominate the thinking
of many scientists in this area.

The marriage of the Edwards model and de Gennes observation brought an
onslaught of several field theoretical methods to derive various scaling
laws describing the static properties of dilute and semi-dilute polymers
solutions. The Edwards model was also generalized to poor solvent conditions
so that polymer collapse could be described. These developments are
summarized in a beautiful monograph by des Cloizeaux and Jannink [9]. 
It is fair
to say that these two landmarks in polymer physics have enabled us to
understand many structural aspects of polymers in solution.

There still are challenges which have come about  in  extending  the Edwards
model to tethered membranes (D= 2) [3]. The demonstration of the
renormalizability of the resulting model is a topic of current research [10]. In
this context there does not appear to be an equivalent spin model which
describes self-avoidance in such objects. Further extension of these
models to membranes and charged species is expected to be important problems
in the general area of soft condensed matter physics. A perusal of the
literature on these topics is suffice to appreciate the deep influence of
the two landmark papers [4,8] in polymer physics.

\textbf{Acknowledgements:} I am grateful to M. E. Fisher and H. Orland for
useful discussions.

\vspace{0.5in}

\textbf{References}

[1] P. J. Flory, J. Chem. Phys. \bf{17}\rm, 303 (1949).

[2a] M. E. Fisher, J. Phys. Soc. Japan \bf{26}\rm (Suppl.), 44 (1969); 
[2b] P. G. de Gennes in "Scaling Concepts in Polymer Physics",
(Cornell University Press, Ithaca, New York, 1985).

[3] Y. Kantor, M. Kardar, and D. R. Nelson, Phys. Rev. A, \bf{35}\rm, 
3056 (1987).

[4] S. F. Edwards, Proc. Phys. Soc. \bf{85}\rm, 613 (1965).

[5] K. F. Freed, Adv. Chem. Phys. \bf{22}\rm, 1 (1972).

[6] C. Domb, J. Chem. Phys., \bf{38}\rm, 2957 (1963).

[7] M. E. Fisher, J. Chem. Phys., \bf{44}\rm, 616 (1966).

[8] P. G. de Gennes, Phys. Lett., \bf{38A}\rm, 229 (1972).

[9] J. des Cloizeaux and G. Jannink in "Polymers in Solution",
(Oxford University Press).

[10] F. David and K. J. Wiese, Nuc. Phys. B, \bf{535}\rm, 555 (1998).

\end{document}